# Determining The Cosmological Constant Using Gravitational Wave Observations

Thomas L. Wilson
*NASA, Johnson Space Center - Houston, Texas USA*
Email: Thomas.Wilson@cern.ch
(November 25, 2019)

It is shown in Einstein gravity that the cosmological constant $\Lambda$ introduces a graviton mass $m_g$ into the theory, a result that will be derived from the Regge-Wheeler-Zerilli problem for a particle falling onto a Kottler-Schwarzschild mass with $\Lambda \neq 0$. The value of $m_g$ is precisely the Spin-2 gauge line appearing on the $\Lambda$-$m_g^2$ phase diagram for Spin-2, the partially massless gauge lines introduced by Deser & Waldron in the $(m_g^2, \Lambda)$ phase plane and described as the Higuchi bound $m_g^2 = 2\Lambda/3$. Note that this graviton is unitary with only four polarization degrees of freedom (helicities $\pm 2$, $\pm 1$, but not 0 because a scalar gauge symmetry removes it). The conclusion is drawn that Einstein gravity (EG, $\Lambda \neq 0$) is a partially massless gravitation theory. Given the recent results measuring the Hubble constant $H_o$ from LIGO-Virgo data, it is then shown that $\Lambda$ can be determined from the LIGO results for the graviton mass $m_g$ and $H_o$. This is yet another multi-messenger source for determining the three parameters $\Lambda$, $m_g$, and $H_o$ in astrophysics and cosmology.

*Introduction:* In order to determine the graviton mass of Einstein gravity (EG), we proceed as follows. A curved Kottler-Schwarzschild (KS) metric with $\Lambda \neq 0$ will be applied to the Regge-Wheeler-Zerilli (RWZ) problem [1-5] representing gravitational radiation perturbations produced by a particle falling onto a large mass M. The RWZ result ($\Lambda=0$) will be extended to the general EG problem with $\Lambda \neq 0$ (EG$\Lambda$), in the fashion that Kottler extended the Schwarzschild metric to de Sitter space (SdS).

One begins with a small perturbative expansion of the Einstein field equations

$$R_{\mu\nu} - \tfrac{1}{2} g_{\mu\nu} R + \Lambda g_{\mu\nu} = -\kappa T_{\mu\nu} \qquad (1)$$

about the known exact solution $\eta_{\mu\nu}$ where the metric tensor is $g_{\mu\nu} = \eta_{\mu\nu} + h_{\mu\nu}$, with $h_{\mu\nu}$ the dynamic perturbation of the background raising and lowering operator $\eta_{\mu\nu}$. The most general spherically symmetric solution is well-known to be a Kottler-Schwarzschild (KS) metric

$$ds^2 = -e^{\nu} dt^2 + e^{\zeta} dr^2 + r^2 d\Omega^2 , \qquad (2)$$

where

$$e^{\nu} = 1 - \frac{2M}{r} - \frac{\Lambda}{3} r^2 = e^{-\zeta} , \qquad (3)$$

with $M = GM^*/c^2$, $d\Omega^2 = (d\Theta + \sin^2 d\phi^2)$, and $\eta_{\mu\nu} = \mathrm{diag}(e^{\nu}, e^{-\nu}, r^2, r^2 \sin 2\Theta)$ in spherically symmetric coordinates. Its contravariant inverse $\eta^{\mu\nu}$ is defined such that $\eta_{\mu\nu} \eta^{\mu\nu} = \delta_\mu^{\ \nu}$.

The wave equation for gravitational radiation $h_{\mu\nu}$ on the non-flat background containing $\Lambda$ in (1) will follow as (9) below, derived now from the procedure developed in the RWZ formalism. Perturbation analysis of (1) for a stable background $\eta^{\mu\nu} = g^{(0)}_{\ \mu\nu}$ produces the following

$$[h_{\mu\nu,\alpha}^{\ ;\alpha} - h_{\mu\alpha,\nu}^{\ ;\alpha} - h_{\nu\alpha;\mu}^{\ ;\alpha} + h_{\alpha\ ;\mu;\nu}^{\ \alpha}] + \eta_{\mu\nu}[h_{\alpha\gamma}^{\ ;\alpha;\gamma} - h_{\alpha\ ;\gamma}^{\ \alpha\ ;\gamma}]$$
$$+ h_{\mu\nu}(R - 2\Lambda) - \eta_{\mu\nu} h_{\alpha\beta} R^{\alpha\beta} = -2\kappa \delta T_{\mu\nu} . \qquad (4)$$

Stability must be assumed in order that $\delta T_{\mu\nu}$ is small. This equation can be simplified by defining the function (introduced by Einstein himself)

$$\bar{h}_{\mu\nu} \equiv h_{\mu\nu} - \tfrac{1}{2} \eta_{\mu\nu} h \qquad (5)$$

and its divergence

$$f_\mu \equiv \bar{h}_{\mu\ \ ;\nu}^{\ \nu} . \qquad (6)$$

Substituting (5) and (6) into (4) and re-grouping terms gives

$$\bar{h}_{\mu\nu,\alpha}^{\ ;\alpha} - (f_{\mu;\nu} + f_{\nu;\mu}) + \eta_{\mu\nu} f_\alpha^{\ ;\alpha} - 2\bar{h}_{\alpha\beta} R^{\alpha\ \beta}_{\ \mu\ \nu} - \bar{h}_{\mu\alpha} R^{\alpha}_{\ \nu} - \bar{h}_{\nu\alpha} R^{\alpha}_{\ \mu}$$
$$+ h_{\mu\nu}(R - 2\Lambda) - \eta_{\mu\nu} h_{\alpha\beta} R^{\alpha\beta} = -2\kappa \delta T_{\mu\nu} . \qquad (7)$$

Now impose the Hilbert-Einstein-de Donder gauge which sets (6) to zero ($f_\mu = 0$), and suppresses any vector gravitons. Wave equation (7) reduces to

$$\bar{h}_{\mu\nu,\alpha}^{\ ;\alpha} - 2\bar{h}_{\alpha\beta} R^{\alpha\ \beta}_{\ \mu\ \nu} - \bar{h}_{\mu\alpha} R^{\alpha}_{\ \nu} - \bar{h}_{\nu\alpha} R^{\alpha}_{\ \mu}$$
$$- \eta_{\mu\nu} h_{\alpha\beta} R^{\alpha\beta} + h_{\mu\nu}(R - 2\Lambda) = -2\kappa \delta T_{\mu\nu} . \qquad (8)$$

In an empty ($T_{\mu\nu} = 0$), Ricci-flat ($R_{\mu\nu} = 0$) space without $\Lambda$ ($R = 4\Lambda = 0$), (8) further reduces to

$$\bar{h}_{\mu\nu,\alpha}^{\ ;\alpha} - 2 R^{\alpha\ \beta}_{\ \mu\ \nu} \bar{h}_{\alpha\beta} = -2\kappa \delta T_{\mu\nu} , \qquad (9)$$

which is the starting point for the RWZ formalism.

*Weak-Field Limit, de Sitter Metric.* The Schwarzschild character of the RWZ problem above will now be relaxed, with $\eta_{\mu\nu}$ again diagonal, but $M = 0$ and $\Lambda \neq 0$ in (2) and (3). The wave equation of paramount importance will follow as

(17).

We know that the trace of the field equations (1) gives $4\Lambda - R = -\kappa T$, whereby they become

$$R_{\mu\nu} - \Lambda g_{\mu\nu} = -\kappa[T_{\mu\nu} - \tfrac{1}{2}g_{\mu\nu}T] \quad . \tag{10}$$

For an empty space ($T_{\mu\nu} = 0$ and $T = 0$), (10) reduces to de Sitter space

$$R_{\mu\nu} = \Lambda g_{\mu\nu} \quad , \tag{11}$$

and the trace to $R = 4\Lambda$.

Substitution of $R$ and $R_{\mu\nu}$ from (11) into (8) using (5) shows that the contributions due to $\Lambda \neq 0$ are of second order in $h_{\mu\nu}$. Neglecting these terms (particularly if $\Lambda$ is very, very small) simplifies (8) to

$$\bar{h}_{\mu\nu,\alpha}{}^{;\alpha} - 2R^{\alpha}{}_{\mu\nu}{}^{\beta}\bar{h}_{\alpha\beta} = -2\kappa\delta T_{\mu\nu} \quad . \tag{12}$$

One can arrive at (12) to first order in $h_{\mu\nu}$ by using $g_{\mu\nu}$ as a raising and lowering operator rather than the background $\eta_{\mu\nu}$ − a result which incorrectly leads some to the conclusion that $\Lambda$ terms cancel in the gravitational wave equation.

Note with caution that (12) and the RWZ equation (9) are not the same wave equation. Overtly, the cosmological terms have vanished from (12), just like (9) where $\Lambda$ was assumed in the RWZ problem to be nonexistent in the first place. However, the character of the Riemann tensor $R^{\alpha}{}_{\mu\nu}{}^{\beta}$ is significantly different in these two relations.

Simplifying the SdS metric by setting the central mass $M^*$ in $\eta_{\mu\nu}$ to zero, produces the de Sitter space (11) of constant curvature $K = 1/R^2$, where we can focus on the effect of $\Lambda$. The Riemann tensor is now

$$R_{\eta\mu\nu\delta} = +K(g_{\gamma\nu}g_{\mu\delta} - g_{\gamma\delta}g_{\mu\nu}) \quad , \tag{13}$$

and reverts to

$$R^{\alpha}{}_{\mu\nu}{}^{\beta} = +K(g^{\alpha}{}_{\nu}g_{\mu}{}^{\beta} - g^{\alpha\beta}g_{\mu\nu}) \quad , \tag{14}$$

for use in (12). This substitution (raising and lowering with $\eta_{\mu\nu}$) into (12) next gives $K$ and $\Lambda$ term contributions

$$-2K[(\bar{h}_{\mu\nu} - \eta_{\mu\nu}\bar{h}) + (\bar{h}_{\alpha\mu}h^{\alpha}{}_{\nu} + \bar{h}_{\nu\beta}h^{\beta}{}_{\mu} - \bar{h}h_{\mu\nu} - \eta_{\mu\nu}h^{\alpha\beta}\bar{h}_{\alpha\beta})]$$
$$+ [2h_{\mu\alpha}\bar{h}^{\alpha}{}_{\nu} + \eta_{\mu\nu}h^2_{\alpha\beta}] \quad , \tag{15}$$

to second order in $h_{\mu\nu}$. Recalling that curvature $K$ is related to $\Lambda$ by $K = \Lambda/3$, substitution of (15) back into (12) gives to first order

$$\bar{h}_{\mu\nu,\alpha}{}^{;\alpha} - \tfrac{2}{3}M\bar{h}_{\mu\nu} + \tfrac{2}{3}\Lambda\eta_{\mu\nu}\bar{h} = -2\kappa\delta T_{\mu\nu} \quad . \tag{16}$$

There is no cancellation of the $\Lambda$ contributions to first order. Noting from (5) that $\bar{h} = h(1-\tfrac{1}{2}\eta)$, then a traceless gauge $\bar{h} = 0$ means either that $h = 0$ or $\eta = 2$. Since $\eta = 4$, (16) reduces to

$$\bar{h}_{\mu\nu,\alpha}{}^{;\alpha} - \tfrac{2}{3}\Lambda\bar{h}_{\mu\nu} = -2\kappa\delta T_{\mu\nu} \tag{17}$$

in a traceless Hilbert-Einstein-de Donder gauge where $\bar{h}_{\mu\nu}{}^{;\nu} = 0$ and $\bar{h}_{\mu}{}^{\mu} = 0$. (17) is a wave equation involving the Laplace-Beltrami operator term $\bar{h}_{\mu\nu,\alpha}{}^{;\alpha}$ for the Spin-2 gravitational perturbation $h_{\mu\nu}$ bearing a mass

$$m_g = \sqrt{2\Lambda/3} \quad , \tag{18}$$

similar to the Klein-Gordon equation $(\Box - m^2)\varphi = 0$ for a Spin-0 scalar field $\varphi$ in flat Minkowski space. The *Locally Flat Limit* section which follows demonstrates that $\bar{h}_{\mu\nu,\alpha}{}^{;\alpha} \to \Box\bar{h}_{\mu\nu}$ in (17) for the limit $r\to 0$. From (17) and (18) then

$$(\Box - m_g^2)\bar{h}_{\mu\nu} = -2\kappa\delta T_{\mu\nu} \tag{19}$$

in the locally flat-space limit $r<<1$.

Note that Penrose [6] has pointed out that due to conformal invariance arguments, the massless Klein-Gordon equation becomes $(\Box - R/6)\varphi = 0$ on a curved background. This necessarily gives (18) since $R=4\Lambda$ in de Sitter space. Also in passing, by rescaling $\bar{h}$ as $\bar{h}_2 \to \tfrac{1}{2}\bar{h}_1$ in (12) and (17), then (18) becomes

$$m_g = \sqrt{\Lambda/3} \quad , \tag{20}$$

which is the surface gravity $\kappa_C = m_g$ of the cosmological event horizon identified by Gibbons & Hawking [7]. It is also found in Weinberg [8].

*Locally Flat Limit of Wave Equation (17).* It is necessary to demonstrate that hidden $\Lambda$-terms arising from $\bar{h}_{\mu\nu,\alpha}{}^{;\alpha}$ in (17) do not cancel the mass term in (18)-(20) when $r\to 0$ and $\bar{h}_{\mu\nu,\alpha}{}^{;\alpha} \to \bar{h}_{\mu\nu,\alpha}{}^{,\alpha} = \Box\bar{h}_{\mu\nu}$, the d'Alembertian in a locally flat region of dS studied above. $\Lambda$-terms appear but cancel out as shown below.

To simplify calculations, now note that $r^2 d\Omega^2$ in (2) is of second-order in $r$ and is negligible as $r\to 0$. Thus the focus is on $e^\nu$ (with $M=0$) in (3) appearing in the diagonal of $\eta_{\mu\nu}$ and its inverse $\eta^{\mu\nu}$. Hence, $\eta_{00}=-c$ and $\eta^{00}=-c^{-1}$, while $\eta_{11}=c^{-1}$ and $\eta^{11}=c$. Also, note that $c(r)\to 1$ and $c(r)^{-1}\to 1$ as $r\to 0$.

Introducing the Christoffel symbol $\Gamma^{\gamma}_{\alpha\beta}$, we can write

$$\bar{h}_{\mu\nu,\alpha}{}^{;\alpha} = g^{\alpha\beta}\bar{h}_{\mu\nu,\alpha;\beta} = g^{\alpha\beta}[\,\bar{h}_{\mu\nu,\alpha;\beta} - (\Gamma^{\varepsilon}_{\alpha\mu}\bar{h}_{\varepsilon\nu})_{,\beta} - (\Gamma^{\varepsilon}_{\alpha\nu}\bar{h}_{\mu\varepsilon})_{,\beta}\,]. \tag{21}$$

Define

$$\bar{h}_{\mu\nu,\alpha}{}^{;\alpha} = \Box\bar{h}_{\mu\nu} + A_{\mu\nu} + B_{\mu\nu} + C_{\mu\nu} \quad , \tag{22}$$

where

$$\Box\bar{h}_{\mu\nu} = \bar{h}_{\mu\nu,\alpha}{}^{,\alpha} \tag{23}$$

$$A_{\mu\nu} = -\Gamma^{\varepsilon}_{\beta\mu}\bar{h}_{\varepsilon\nu}{}^{,\beta} - \Gamma^{\varepsilon}_{\beta\nu}\bar{h}_{\varepsilon\mu}{}^{,\beta} - \Gamma^{\varepsilon}_{\beta\alpha}\bar{h}_{\mu\nu,\varepsilon}\eta^{\alpha\beta}$$
$$-\Gamma^{\varepsilon}_{\alpha\mu}\bar{h}_{\varepsilon\nu}{}^{,\alpha} - \Gamma^{\varepsilon}_{\alpha\nu}\bar{h}_{\mu\varepsilon}{}^{,\alpha} \tag{24}$$

$$B_{\mu\nu} = -(\Gamma^{\varepsilon}_{\alpha\mu})^{,\alpha}\bar{h}_{\varepsilon\nu} - (\Gamma^{\varepsilon}_{\alpha\nu})^{,\alpha}\bar{h}_{\mu\varepsilon} \tag{25}$$

$$C_{\mu\nu} = -\eta^{\alpha\beta}[(\Gamma^{\varepsilon}_{\beta\delta}\Gamma^{\delta}_{\alpha\mu} - \Gamma^{\varepsilon}_{\beta\alpha}\Gamma^{\delta}_{\delta\mu} - \Gamma^{\varepsilon}_{\beta\mu}\Gamma^{\delta}_{\alpha\delta})\bar{h}_{\varepsilon\nu}$$
$$-\Gamma^{\delta}_{\beta\varepsilon}\Gamma^{\varepsilon}_{\alpha\mu}\bar{h}_{\delta\nu} - \Gamma^{\delta}_{\beta\nu}\Gamma^{\varepsilon}_{\alpha\mu}\bar{h}_{\varepsilon\delta}$$
$$+(\Gamma^{\varepsilon}_{\beta\delta}\Gamma^{\delta}_{\alpha\nu} - \Gamma^{\varepsilon}_{\beta\alpha}\Gamma^{\delta}_{\delta\nu} - \Gamma^{\varepsilon}_{\beta\nu}\Gamma^{\delta}_{\alpha\delta})\bar{h}_{\mu\varepsilon}$$
$$-\Gamma^{\delta}_{\beta\mu}\Gamma^{\varepsilon}_{\alpha\nu}\bar{h}_{\delta\varepsilon} - \Gamma^{\delta}_{\beta\varepsilon}\Gamma^{\varepsilon}_{\alpha\nu}\bar{h}_{\mu\delta}\,] \quad . \tag{26}$$

$B_{\mu\nu}$ is the term of interest. $A_{\mu\nu}$ and $C_{\mu\nu}$ contain factors of second order, or terms that vanish in locally flat space ($r<<1$). Furthermore, only the first-order second derivatives in $B_{\mu\nu}$ remain as $r\to 0$. These terms are

$$B^*_{\alpha\mu\nu}{}^{,\alpha} = -\tfrac{1}{2}\eta^{\varepsilon\gamma}[\,(\eta_{\alpha\gamma,\mu}{}^{,\alpha} + \eta_{\mu\gamma,\alpha}{}^{,\alpha} - \eta_{\alpha\mu,\gamma}{}^{,\alpha})\bar{h}_{\varepsilon\nu}$$
$$+ (\eta_{\alpha\gamma,\nu}{}^{,\alpha} + \eta_{\nu\gamma,\alpha}{}^{,\alpha} - \eta_{\alpha\nu,\gamma}{}^{,\alpha})\bar{h}_{\mu\varepsilon}\,] \tag{27}$$

which can be defined as

$$B^{*\ \alpha}_{\alpha\mu\nu} = F_{\mu\nu} + G_{\mu\nu} + H_{\mu\nu} , \qquad (28)$$

where

$$F_{\mu\nu} = -\frac{1}{2}\eta^{\varepsilon\gamma}\left[(\Box\eta_{\mu\gamma})\bar{h}_{\varepsilon\nu} + (\Box\eta_{\nu\gamma})\bar{h}_{\mu\varepsilon}\right] \qquad (29)$$

$$G_{\mu\nu} = -\frac{1}{2}\eta^{\varepsilon\gamma}\left[\eta_{\alpha\gamma,\mu}{}^{,\alpha}\bar{h}_{\varepsilon\nu} + \eta_{\alpha\gamma,\nu}{}^{,\alpha}\bar{h}_{\mu\varepsilon}\right] \qquad (30)$$

$$H_{\mu\nu} = +\frac{1}{2}\eta^{\varepsilon\gamma}\left[\eta_{\alpha\mu,\gamma}{}^{,\alpha}\bar{h}_{\varepsilon\nu} + \eta_{\alpha\nu,\gamma}{}^{,\alpha}\bar{h}_{\mu\varepsilon}\right] . \qquad (31)$$

In this approximation, $\Box = -\partial_t^2 + \nabla^2 \to \nabla^2$. Also

$$\Box\eta_{00} \to \nabla^2\eta_{00} = +\tfrac{2}{3}\lambda \text{ and } \Box\eta_{11} \to \nabla^2\eta_{11} = +\tfrac{2}{3}\lambda.$$

We find that

$$F_{\mu\nu} = -\frac{1}{2}\eta^{00}\left[(\Box\eta_{\mu 0})\bar{h}_{0\nu} + (\Box\eta_{\nu 0})\bar{h}_{\mu 0}\right]$$
$$-\frac{1}{2}\eta^{11}\left[(\Box\eta_{\mu 1})\bar{h}_{1\nu} + (\Box\eta_{\nu 1})\bar{h}_{\mu 1}\right] \qquad (32)$$

whereby (all other terms do not contribute)

$$F_{00} = -\eta^{00}\left[(\Box\eta_{00})\bar{h}_{00}\right] = +\tfrac{2}{3}\lambda\bar{h}_{00} \qquad (33)$$

$$F_{11} = -\eta^{11}\left[(\Box\eta_{11})\bar{h}_{11}\right] = -\tfrac{2}{3}\lambda\bar{h}_{11} . \qquad (34)$$

Next

$$G_{\mu\nu} = -\frac{1}{2}\eta^{11}\left[\eta_{11,\mu}{}^{,1}\bar{h}_{1\nu} + \eta_{11,\nu}{}^{,1}\bar{h}_{\mu 1}\right] \qquad (35)$$

whereby (all other terms do not contribute)

$$G_{01} = -\tfrac{1}{3}\lambda\bar{h}_{01}; \quad G_{10} = -\tfrac{1}{3}\lambda\bar{h}_{10}; \quad G_{11} = -\tfrac{2}{3}\lambda\bar{h}_{11} . \qquad (36)$$

And lastly,

$$H_{\mu\nu} = \frac{1}{2}\eta^{11}\left[\eta_{\alpha\mu,1}{}^{,\alpha}\bar{h}_{1\nu} + \eta_{\alpha\nu,1}{}^{,\alpha}\bar{h}_{\mu 1}\right] , \qquad (37)$$

whereby

$$H_{00} = 0 ; \quad H_{11} = \tfrac{2}{3}\lambda\bar{h}_{11} ;$$
$$H_{01} = \tfrac{1}{3}\lambda\bar{h}_{01} ; \quad H_{10} = \tfrac{1}{3}\lambda\bar{h}_{10}. \qquad (38)$$

Summarizing, the two contributing terms to $F_{\mu\nu}$ in (33) and (34) are equal and opposite thereby cancelling in (32). Thus, $F_{\mu\nu}=0$. Similarly, the collective $G_{\mu\nu}$ and $H_{\mu\nu}$ terms in (36) and (38) cancel one another, giving $G_{\mu\nu} + H_{\mu\nu} = 0$. Hence $B^{*\ \alpha}_{\alpha\mu\nu} = B_{\mu\nu} \equiv 0$ in (28) and (25). Therefore we get $\bar{h}_{\mu\nu,\alpha}{}^{;\alpha} \to \bar{h}_{\mu\nu,\alpha}{}^{,\alpha} = \Box\bar{h}_{\mu\nu}$ in the locally flat limit of (17).

The graviton mass (18) for EGΛ thus follows from this analysis, a result first determined many years ago [9].

*Identifying Einstein Gravity As A Partially Massless Theory.* The cosmological phase diagrams for partially massless fields of arbitrary spin in de Sitter space (Λ≠0) are well understood thanks to the seminal work of Deser & Nepomechie [10] and Deser & Waldron [11-17], in conjunction with that of Higuchi [18-21].

(18) removes the scalar helicity-0 mode along the Higuchi partially-massless gauge line for Spin-2, leaving only 4 instead of 5 propagating degrees of freedom [15] – hence the term partially massless gravity. With respect to gravitational wave polarization analysis, this partially massless feature of EGΛ went unnoticed earlier on in initial polarization studies of gravitational waves which focused on Pauli-Fierz massive gravity effects [22-25]. The latter do not address partial masslessness in gravitational radiation behavior.

Derived directly from EGΛ in (1)-(3), (18) proves that EGΛ is a partially massless theory because that is specifically the Higuchi bound established by Deser and Nepomechie [10], Deser and Waldron [11-17], and articulated by Higuchi [18-21]. Massive gravity thus finds its roots when Einstein first introduced Λ into GR, rather than later when Pauli & Fierz (P-F) [26] pursued the study of massive gravity by adding appropriate terms to the Einstein-Hilbert Lagrangian.

*Determining Λ From Gravitational Wave Observations.* (18) is hence a direct prediction of EGΛ in (1). Recalling that gravitational wave observations can be used to determine the Hubble constant $H_o$ [27, 28], we know that $H_o^2 = \Lambda/3$ in de Sitter space [8, Eq. 2.6] from which Λ can be determined. Given the currently known disparity in $H_o$ determinations [29, 30], Λ, $m_g$, and $H_o$ must eventually be brought into reconciliation. The question now becomes how to measure these effects using LIGO, VIRGO, and future LISA antenna configurations to determine whether polarization measurements can establish the loss of the helicity 0 excitation due to a scalar gauge symmetry but not the loss of helicity ±1, as predicted by the partially massless theory [12, 31].

*In Conclusion.* These results come directly from the RWZ equation (9). The consequence is yet another way to determine the cosmological constant Λ, but from gravitational wave observations. It constitutes an entirely new prediction from Einstein's theory, that Λ, $c$, $H_o$, and $m_g$ (having only 4 Spin-2 DOFs with helicities ±2, ±1), and conventional Λ-lore such as dark matter in ΛCDM models, are interrelated. For that reason alone, (18) needs to be verified experimentally. In addition, all of these parameters must collectively produce self-consistent values. The answer may also contribute to our understanding of galactic-rotation-curve behavior and the accelerating Universe should the much-discussed Yukawa-potential implications of an $m_g$ like (18) prove to be true. Such predictions by EGΛ need to be investigated further.

The fundamental question for partially massive gravity is whether existing gravitational wave antenna configurations can be used to measure or determine the loss of the helicity 0 polarization caused by loss of a scalar gauge symmetry. It will probably require additional antenna configurations and possibly more antennas.

———————

[1] T. Regge, T. and J.A. Wheeler, Phys. Rev. **108**, 1063 (1957).
[2] P.C. Peters, Phys. Rev. **146**, 938 (1966).
[3] R.A. Isaacson, Phys. Rev. **166**, 1263 (1968); *ibid*., 1272.
[4] F.J. Zerilli, Phys. Rev. **D2**, 2141 (1970).
[5] F.J. Zerilli, Phys. Rev. Lett. **24**, 737 (1970).
[6] Penrose, R., in *Relativity, Groups, & Topology*, C. & B. DeWitt (Gordon & Breach, 1964) 565-584.

[7] G.W. Gibbons and S. Hawking, Phys. Rev. **D15**, 2738 (1977).
[8] S. Weinberg, Rev. Mod. Phys. **61**, 1 (1989).
[9] T.L. Wilson, "Gravitational Radiation Theory," Master's Thesis, Rice University (1973), Houston, App. Y. Published as NASA Tech. Memorandum (1973). Online at NASA Tech Reports Server (NTRS): NASA TMX-58132.
[10] S. Deser and R.I. Nepomechie, Phys. Lett. **B132**, 321 (1983); Annals Phys. **154**, 396 (1984).
[11] S. Deser and A. Waldron, Phys. Rev. Lett. **87**, 031601 (2001).
[12] S. Deser and A. Waldron, Phys. Lett. **B508**, 347 (2001).
[13] S. Deser and A. Waldron, Phys. Lett. **B513**, 137 (2001).
[14] S. Deser and A. Waldron, Nucl. Phys. **B607**, 577 (2001).
[15] S. Deser and A. Waldron, Nucl. Phys. **B631**, 369 (2002).
[16] S. Deser and A. Waldron, Nucl. Phys. **B662**, 379 (2003).
[17] S. Deser and A. Waldron, Phys. Lett. **B603**, 30 (2004).
[18] A. Higuchi, Nucl. Phys. **B282**, 397 (1987)
[19] A. Higuchi, Nucl. Phys. **B325**, 745 (1989).
[20] A. Higuchi, J. Math. Phys. **28**, 1553 (1987).
[21] A. Higuchi, Nucl. Phys. **B282**, 397 (1987).
[22] Eardley, D.M., *et al.*, Phys. Rev. Lett. **30**, 884 (1973).
[23] Will, C.M., Liv. Rev. Rel. **17**, 4 (2014).
[24] Will, C.M., Liv. Rev. Rel. **9**, 3 (2006).
[25] Abbott, B.P., et al., Phys. Rev. Lett. **120**, 201102 (2018).
[26] M. Fierz and W. Pauli, Proc. Roy. Soc. Lond. **A173**, 211 (1939).
[27] B.F. Schutz, Nature **323**, 310 (1986).
[28] LIGO-VIRGO Collaborations, Nature **551**, 85.
[29] V. Poulin *et al*., Phys. Rev. Lett. **122** (2019) 221301.
[30] A.G. Riess *et al.*, Ap. J. 876 (2019) 85.
[31] S. Deser, Int'l.J.Mod.Phys., A17(2002)32, ArXiv_0110027.